\documentclass[twocolumn,showpacs,prc]{revtex4}          
\newcommand{ \be }{\begin{eqnarray}}
\newcommand{ \ee }{\end{eqnarray}}
\newcommand{ \la }{\langle}
\newcommand{ \ra }{\rangle}

\newcommand{ \bv }{{\bf v}}
\newcommand{ \bB }{{\bf B}}
\newcommand{ \br }{{\bf r}}
\newcommand{ \bR }{{\bf R}}
\newcommand{ \eps }{\varepsilon}
\newcommand{ \mean }[1]{\left\langle #1 \right\rangle}

\def\P{$\cal P$}
\def\CP{$\cal CP$}

\newcommand{ \psirp }{\Psi_{RP}}
\newcommand{ \phia }{\phi_{\alpha}}
\newcommand{ \phib }{\phi_{\beta}}

\usepackage{color}

\usepackage{graphicx} % Include figure files
\usepackage{dcolumn} % Align table columns on decimal point          
\usepackage{bm} % bold math          

\begin{document}          
\title{       
%\begin{flushright}  \small \sl version 00,  \today \\  \end{flushright} 
Testing the Chiral Magnetic Effect with Central $U+U$ collisions
} 
\author{Sergei A. Voloshin}
\address{Wayne State University, Detroit, Michigan 48201, USA}

\begin{abstract}     
A quark interaction with topologically nontrivial gluonic fields,
instantons and sphalerons, violates \P~ and \CP~ symmetry.
In the strong magnetic field of a noncentral nuclear collision
such interactions lead to the charge separation along the magnetic
field, the so-called chiral magnetic effect (CME).
Recent results from the STAR collaboration on charge dependent 
correlations are consistent with theoretical expectations for CME
 but may have contributions from other effects, which prevents
definitive interpretation of the data. 
Here I propose to use central body-body $U+U$ collisions 
to disentangle correlations due to CME from
possible background correlations due to elliptic flow.
Further more quantitative studies can be performed with
collision of isobaric beams.
\end{abstract} 
 
\pacs{25.75.Ld}          
  
\maketitle

\section{Introduction}  
Quantum chromodynamics (QCD), the theory of strong interactions, 
is a non-Abelian gauge theory that possesses multiple 
vacua characterized by Chern-Simons numbers.
QCD links chiral symmetry breaking and the 
origin of hadron masses to the existence of topologically nontrivial 
classical gluonic fields, instantons and sphalerons, 
describing the transitions between the vacuum states with 
different Chern-Simons numbers.  
Quark interactions with such fields change the quark chirality
and are $\cal P$~and $\cal CP$~odd. 
For a review, see Refs.~\cite{Shuryak:1997vd,Diakonov:2002fq}.
Though theorists have little doubt in the existence of such fields,
they have never been observed directly, e.g. at the level of quarks in
the deep inelastic scattering. 
It was suggested in Ref.~\cite{Kharzeev:1998kz} to look for 
metastable $\cal P$~and $\cal CP$~odd domains,
space-time regions occupied by a classical field with a nonzero topological
charge, in ultrarelativistic heavy ion collision.
Some earlier discussion of how one could observe this
{\em  local strong  parity  violation} 
%and suggested observables 
can be found 
in Refs.~\cite{Kharzeev:1998kz,Voloshin:2000xf,Finch:2001hs}.
 
The situation with experimental search for the local strong parity
violation drastically
changed once it was noticed~\cite{Kharzeev:2004ey,Kharzeev:2007tn} 
that in noncentral nuclear collisions it would lead to
the asymmetry in the emission 
of positively and negatively charged particle perpendicular 
to the reaction plane.
Such charge separation is a consequence of the difference in the number
of quarks with positive and negative helicities positioned
in the strong magnetic field ($\sim 10^{15}$~T) 
of  a noncentral nuclear collision, the so-called 
chiral magnetic effect (CME)~\cite{Kharzeev:2004ey,Kharzeev:2007jp}. 
The same phenomenon can also be understood as an effect of the induced electric
field that is parallel to the static external magnetic field,
{\em chiral magnetic induction}, 
which occurs in the presence of topologically nontrivial 
gluonic fields~\cite{Fukushima:2008xe}.
It has been also argued that the charge separation could have 
origin in nonzero vorticity of the system created 
in noncentral collisions~\cite{Kharzeev:2007tn}.
Chiral magnetic effect has been observed in the lattice QCD
calculations~\cite{Buividovich:2009wi,Abramczyk:2009gb,Buividovich:2009zzb}.
Newer developments in this field has been recently discussed at the RIKEN
BNL  workshop~\cite{workshop}. 
  
An experimental observation of CME would 
be a direct proof for the existence of topologically nontrivial
vacuum structure and would provide an opportunity for a direct
experimental study of the relevant physics.
The difficulty in experimental observation of CME comes from the fact 
that the direction of the charge separation varies 
%event by event 
in accord with the sign of the topological charge of the domain.
Then the observation of the effect is possible only by correlation techniques.
According to Refs.~\cite{Kharzeev:2004ey,Kharzeev:2007tn,Kharzeev:2007jp}
the charge separation could lead to asymmetry in particle production
 $(N_- -N_+)/(N_-+N_+)\sim Q/N_{\pi^+}$, where $Q=0,\pm 1,\pm 2, ...$ 
is the topological charge and
 $N_{\pi^+}$ is the positive pion multiplicity in 
one unit of rapidity 
-- the typical scale of such correlations.
It results in correlations of the order of $10^{-4}$, which is
accessible in current high statistics heavy ion experiments.
An observable directly sensitive to the charge
separation effect has been proposed in Ref.~\cite{Voloshin:2004vk}.
It is discussed in more detail below.

Recent STAR results~\cite{:2009uh,:2009txa}
on charge separation relative to the reaction plane
%are 
consistent with the expectation for CME
can be considered as evidence of the local strong parity violation. 
The ambiguity in the interpretation
of experimental results comes from possible contribution of 
(the reaction plane dependent) correlations not related to CME. 
As the detailed quantitative predictions for CME do not yet
exist, it is difficult to disentangle different contributions. 
A key ingredient to CME is the strong magnetic field, 
while the background effects originate in the elliptic flow.
In noncentral collisions of spherical nuclei
such as gold, both, magnetic field, and the elliptic flow are
strongly (cor)related to each other. 
In order to disentangle the two effects one has to find a possibility 
to significantly change the relative strengths of the magnetic field 
and the elliptic flow.
The discussion of such a possibility provided by 
$U+U$ collisions is the subject of this Letter. 
Necessary quantitative estimates are obtained using
Glauber Monte-Carlo simulations. 
I estimate the magnetic field 
following the approach of Ref.~\cite{Skokov:2009qp}.
Estimates of elliptic flow are based on the assumption that it scales
with initial (participant) eccentricity of the nuclei overlap region. 

\begin{figure}
\centerline{\includegraphics[width=0.32\textwidth]{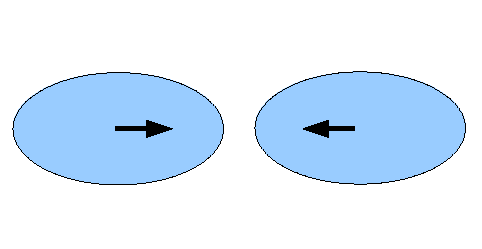} }
	\centerline{(a) }
\centerline{\includegraphics[width=0.2\textwidth]{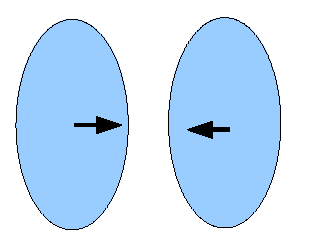} }
	\centerline{(b) }
 \caption{Schematic view of central $U+U$ collisions:
(a) tip-tip   and    (b) body-body.
}
\label{figUU}
\end{figure}

In noncentral nuclear collisions particle distribution in azimuthal
angle is not uniform. The deviation from a flat distribution is
called anisotropic flow and often is described by the Fourier
decomposition~\cite{Voloshin:1994mz,Poskanzer:1998yz} (for a review,
see Ref.~\cite{Voloshin:2008dg}):
\be
 \frac{dN_\alpha}{d\phi} &\propto&  1+ 2 v_{1,\alpha} \cos(\Delta \phi)+
2\, v_{2,\alpha} \cos(2\Delta\phi)+... 
\nonumber \\
&+& 2  a_{1,\alpha} \sin(\Delta \phi) +
2\, a_{2,\alpha} \sin(2 \Delta \phi) +... \, ,
\label{eq:expansion}
\ee
where $\Delta \phi =(\phi-\psirp)$ is the particle azimuth 
relative to the reaction plane,  % (see Fig.~1),
$v_1$ and $v_2$  account for directed and elliptic flow. 
Subscript $\alpha$ 
%is used to 
denotes the particle type.
Because of the ``up-down'' symmetry of the collisions $a_{n}$ coefficients 
are usually omitted.
CME violates such a symmetry. 
Although the  ``direction'' of
the violation fluctuates event to event and on average is zero,
in events with a particular sign of the topological charge, the
average is not zero. 
As a result, it leads to a nonzero contribution to correlations,
One expects that the first harmonic would account for the most of the
effect. 
To measure $\mean{a_{1,\alpha} a_{1,\beta}}$, 
it was proposed~\cite{Voloshin:2004vk} to use the correlator:
\be
\hspace*{-2cm}
& \mean{ \cos(\phia +\phib -2\psirp) } = 
\label{eq:obs1}
\\  
& \mean{\cos\Delta \phia\, \cos\Delta \phib} 
-\mean{\sin\Delta \phia\,\sin\Delta \phib}
\nonumber
\\ 
& =
[\mean{v_{1,\alpha}v_{1,\beta}} + B_{in}] - [\mean{a_{q,\alpha} a_{1,\beta}}
+ B_{out}]
\nonumber 
\\
& \approx - \mean{a_{1,\alpha} a_{1,\beta}} + [ B_{in} - B_{out}].
\label{eq:correlator}
\ee
This correlator represents the difference between 
correlations ``projected'' onto 
the reaction plane and the correlations projected onto an axis
perpendicular to the reaction plane (a more detailed discussion of that
can be found in Refs.~\cite{Voloshin:2004vk,:2009txa}. 
The key advantage of using such a difference is that it
removes the correlations among
particles $\alpha$ and $\beta$ that are not related to the reaction plane 
orientation.
The remaining background in Eq.~\ref{eq:correlator},  $B_{in}
-B_{out}$, is due to processes in which particles $\alpha$ and
$\beta$ are products of a cluster (e.g. resonance, jet,
 di-jets) decay, and the cluster itself exhibits elliptic
flow or decays (fragments)
differently when emitted in-plane compared to out-of-plane.
The corresponding contribution to the correlator can be estimated 
as~\cite{Voloshin:2004vk,:2009txa}:
\be  
&\la \cos(\phi_\alpha + \phi_\beta -2\psirp) \ra =
\label{eq:resonance}
\\
&\frac{ N_{\frac{clust}{event}} N_{\frac{pairs}{clust}}} 
{ N_{\frac{pairs}{event}}}    
          \,
\la \cos(\phi_\alpha + \phi_\beta -2\phi_{clust}) \ra_{clust}
\; v_{2,clust},
\nonumber
\ee 
where $\la ... \ra_{clust}$ indicates that the average is performed only
over pairs consisting of two daughters from the same cluster.
This kind of background can not be easily eliminated or suppressed
and constitute the main uncertainty in the interpretation of the STAR results. 
To address its contribution one has to rely on model 
calculations~\cite{:2009txa}.
A better approach would be to perform experiments where the relative
contribution of CME and background can be varied.
Note that the background is proportional to the elliptic flow present
in the event, $v_{2,clust}$ in Eq.~\ref{eq:resonance}.
It is not clear how one could suppress elliptic flow and at the same 
time preserve strong magnetic field needed for CME. 
But the opposite, collisions with strong elliptic flow and no (or
almost no) magnetic field, seems to be possible. 
This can be achieved in central body-body $U+U$ collisions.  
Uranium nuclei are not spherical and have roughly ellipsoidal shape.
Central collision, when most of the nucleons interact, can have
different geometry, ranging from the so-called tip-tip collisions to
body-body collisions~\cite{Nepali:2007an}, see Fig.~\ref{figUU}. 
Unlike tip-tip collisions, body-body ones would exhibit strong elliptic 
flow. 
Neither would lead to a strong magnetic field; consequently, 
a very weak signal due to CME is expected, while background
would be much stronger in body-body configuration compared to tip-tip
configuration.

Collisions of uranium nuclei were first proposed for RHIC by 
P.~Braun-Munzinger~\cite{pbm} with the goal to achieve higher energy 
density compared to Au+Au collisions. The idea was later 
elaborated in Refs.~\cite{Shuryak:1999by,Kuhlman:2005ts}, in particular 
pointing out an important possibility to study elliptic flow at such high 
energy densities. 

At RHIC one can select central collisions by requiring low signal 
in the zero degree calorimeters that detect spectator neutrons. 
Below I discuss how one might ``control'' the geometry of the collision, 
and, consequently, the relative strengths
of the signal due to CME and background,  by selecting events based on
multiplicity, signal in the zero degree calorimeters, and the magnitude
of the flow vectors.  

The magnetic field strength at a position $\br$ and time $t$ is
defined by the Lienard-Wiechert potentials 
\begin{equation}
e \bB (t, \br) = \alpha_{\rm{EM}} \sum_n e_n  
\frac{(1-v_n^2)}{\left(R_n - \bR_n\bv_n\right)^{3}} 
 \bv_n \times \bR_n , 
\label{Linard}
\end{equation}
where $\alpha_{\rm{EM}}\approx1/137$
is the fine-structure constant, and 
$e_n$ is the electric charge of the $n$th particle in units of the
electron charge. 
$\bR_n = \br-\br_n$, where $\br_n$ is the radius vector of particle, 
$\bv_n$ is particle velocity. 
The quantities $\bv_n$ and $\br_n$ are taken at retarded time 
$t'=t-| \br-\br_n(t') |$.  
Summation runs over all charged spectators.
Spectator contribution to the magnetic field is dominant at early 
times~\cite{Kharzeev:2007jp,Skokov:2009qp}; we also use this approximation
in our estimates.
Because of the Lorentz contraction, in collisions of ultrarelativistic
nuclei,
 the longitudinal size of the nucleus is negligible compared to the
transverse size, and the time dependence of the
magnetic field is totally determined by the gamma-factor (energy)
of colliding nuclei.
We are interested only in a relative change in the strength of 
the magnetic field in collisions at different centralities and different
configurations.  
For this, it is sufficient to calculate the magnetic field only at
$t=0$ (the time the two nuclei collide). 
At $\sqrt{s_{NN}}=200$~GeV, the collision energy used in our estimates,  
the magnetic field at $t=0$ is about factor of 2
smaller compared to the maximum value reached approximately
at $t \approx 0.05$~fm/c~\cite{Skokov:2009qp}.   

Elliptic flow is determined by the geometry of the overlapping
zone. We assume  $v_2=\kappa \eps_p$,
where $\eps_p$ is the so-called participant eccentricity. 
We consider only events within $<5$\% of the most central collisions,
for which  $\kappa \approx const$.
For definitions of eccentricity and details of
experimental measurements of $v_2/\eps$, see
review~\cite{Voloshin:2008dg}.
We take $\kappa=0.2$ based on experimental measurements.
Initial eccentricity, magnetic field, and charged
particle multiplicity at midrapidity are calculated 
using Glauber Monte-Carlo with all
parameters taken the same as used in Ref.~\cite{Filip:2009zz}.

\begin{figure}
\includegraphics[width=0.42\textwidth]{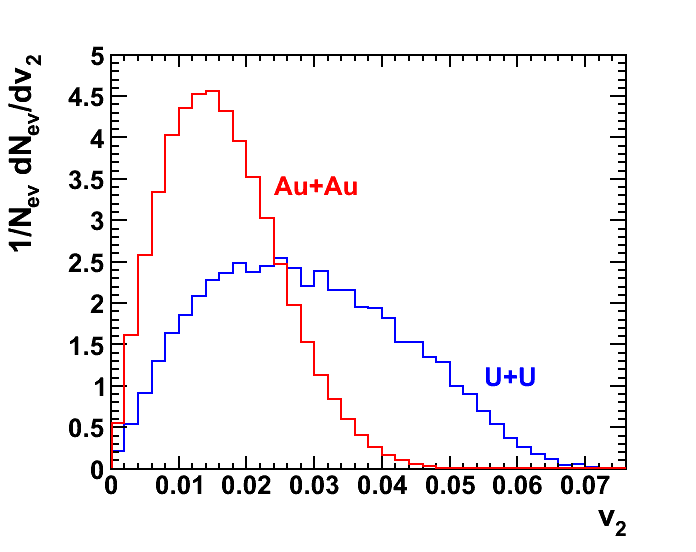}
 \caption{
(Color online)
Event distributions in $v_2$ for Au+Au and $U+U$
   collisions in event samples with the number of spectators $N_{sp}<20$.
}
 \label{fig:ecc}
\end{figure}

\begin{figure}
\includegraphics[width=0.42\textwidth]{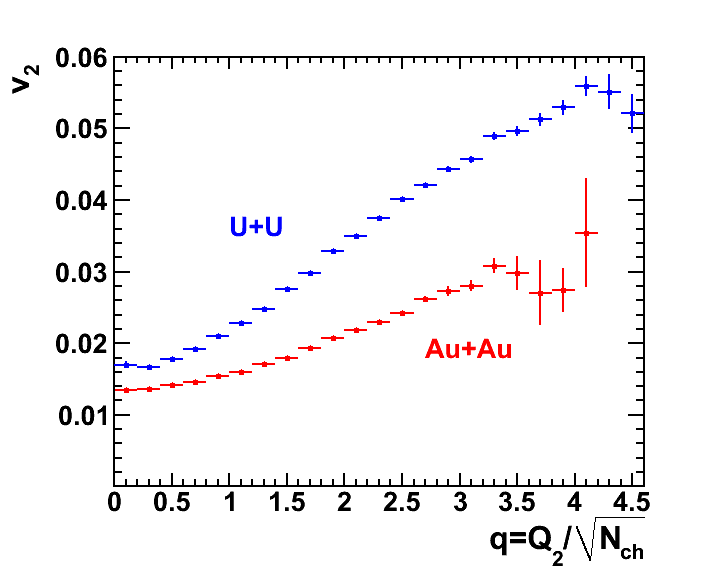}
	\centerline{(a) }
\includegraphics[width=0.42\textwidth]{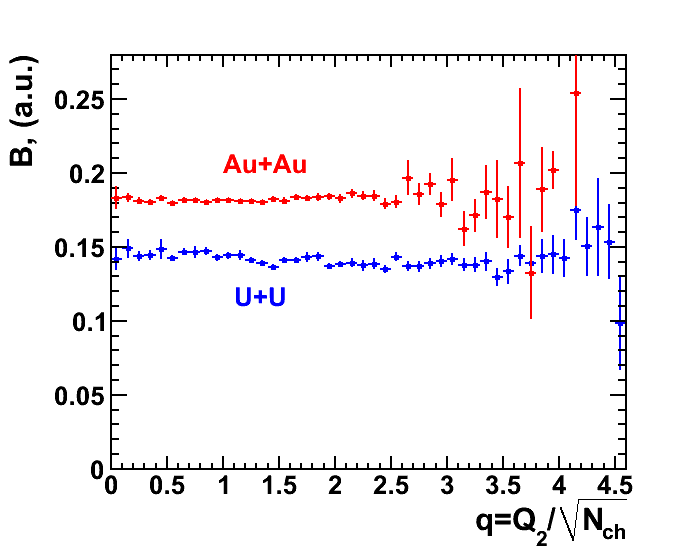}
	\centerline{(b) }
 \caption{
(Color online)
Elliptic flow and the magnetic field (in arbitrary units)
as a function of $q$, the magnitude of the flow vector, in events with 
the number of spectators $N_{sp}<20$.
}
 \label{fig:v2B}
\end{figure}

The effect of nonsphericity of uranium nuclei is clearly visible in
Fig.~\ref{fig:ecc} which shows the distribution of events in elliptic
flow $v_2$ in event samples with number of spectators, $N_{sp}<20$.
The average elliptic flow is almost a factor of 2 larger in $U+U$
collisions compared to Au+Au, which would mean a strong increase 
in the background correlations compared to that of due to CME.
The requirement of the same number of spectators assures that the
magnetic field is not very different in the two systems;
it is slightly lower in $U+U$ collisions [see
Fig.~\ref{fig:v2B}b].
The condition  $N_{sp}<20$ selects about 1.5\% of the most central
events in $U+U$
collisions and about 2.3\% in Au+Au collisions.

Within the event sample selected on the basis of number of spectators,
it would be instructive to study the dependence of the signal 
on the magnitude of the elliptic flow. 
As a measure of the latter we use the
magnitude of the flow vector $q=Q_2/\sqrt{M}$, where 
\be
Q_{2,x}=\sum_{i=1}^M \cos(2\phi_i),\;\;
Q_{2,y}=\sum_{i=1}^M \sin(2\phi_i),
\ee 
and the sum runs over all particles in a given momentum window. 
%In this study 
We calculate the flow vector based on charged particle 
multiplicity in 2 units of rapidity.
As shown in Fig.~\ref{fig:v2B}, the elliptic flow 
is strongly correlated with $q$, and 
at the same time the magnetic field has almost no $q$-dependence.
It means that the correlator, Eq.~\ref{eq:correlator}, used to measure the
signal would 
 stay constant if the signal is mostly determined by CME, and increase
strongly with $q$ if it is due to the background effects.
For such a test $U+U$ collisions provide a significantly better opportunity
than Au+Au collisions. First, the relative variation in $v_2$ is
almost a factor of 2 larger that that in Au+Au collisions. 
Also, the variation in elliptic flow in Au+Au collision is mostly
determined by fluctuations in the initial eccentricity, which are still
not very well known. In $U+U$ collisions the elliptic flow variation is
mostly due to variation in orientation of the nuclei at the moment of
collision. The corresponding estimates have much smaller uncertainty.

\begin{figure}
\includegraphics[width=0.38\textwidth]{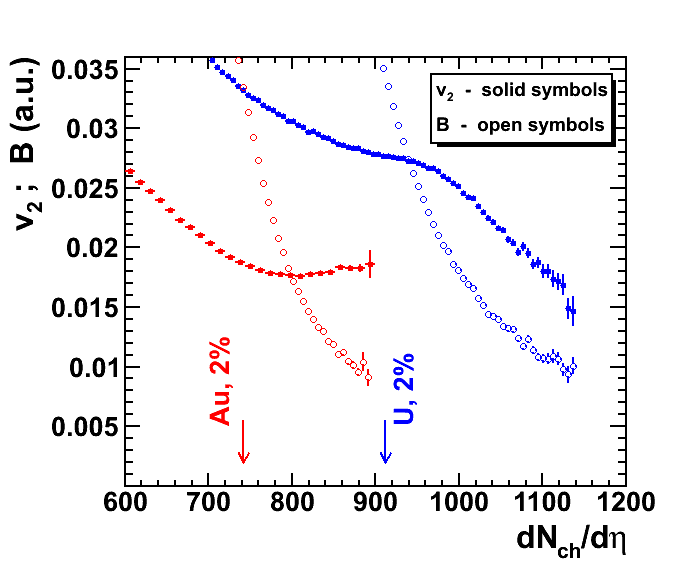}
 \caption{
(Color online)
Elliptic flow and the magnetic field (arbitrary units)
in Au+Au and $U+U$   collisions as function of multiplicity. 
The arrows indicate the
   multiplicities corresponding to the top 2\% of the collision cross section.
}
 \label{fig:Bv2nch}
\end{figure}

While selection of the events based on the number of spectators 
is very useful, it seems to be also possible to
disentangle CME and background correlations based only on the
dependence of the signal on charged multiplicity. 
Figure~\ref{fig:Bv2nch} presents the dependence of the elliptic flow and
magnetic field on charged multiplicity.
The elliptic flow dependence is different for two systems, 
with $U+U$ collisions exhibiting a characteristic kink (cusp) 
at multiplicity $\sim 1000$~\cite{Filip:2009zz}, reflecting the fact that
high(er) multiplicity events have predominantly tip-tip 
orientation;
the latter also leads to a decrease in elliptic flow.
Being mostly determined
by correlation of the multiplicity with the number if participants, 
the magnetic field has similar dependence on multiplicity 
for both collision systems. 
The difference in the dependencies of the
magnetic field and elliptic flow on charged multiplicity can be used a
as a test for the nature of correlations contributing to the signal.

The charge separation dependence on the strength of the magnetic
field can be further studied with collision of isobaric
nuclei, such as  $^{96}_{44}Ru$ and    $^{96}_{40}Zr$.
These nuclei have the same mass number, but differ by the charge.
The multiparticle production 
in the midrapidity region would be affected very
little in collision of such nuclei, and 
one would expect very similar elliptic flow. 
At the same time the magnetic field would be proportional to the
nuclei charge and can vary by more than 10\%, which can results in
20\% variation in the signal.  
Such variations should be readily measurable.
The collisions of  $^{96}_{44}Ru$ and    $^{96}_{40}Zr$
 isotopes have been successfully 
used at GSI~\cite{Hong:2001tm} in a study of baryon stopping.
Collisions of isobaric nuclei at RHIC will be also extremely valuable
for understanding the initial conditions, and in particular the
initial velocity fields, the origin of directed flow, etc.

In summary, the estimates presented in this Letter show that a detailed
analysis of central Au+Au and $U+U$ collisions should be able to
disentangle CME and background correlations contributing to the signal
observed by STAR. 

Discussions with J.~Dunlop and P.~Filip are gratefully acknowledged.
This work was supported in part by the US Department of Energy, 
Grant No. DE-FG02-92ER40713.

%==========================================================================
%==========================================================================  
  
\end{document}